\documentclass{article}

\usepackage{arxiv}

\usepackage[utf8]{inputenc} 
\usepackage[T1]{fontenc}    
\usepackage{hyperref}       
\usepackage{url}            
\usepackage{booktabs}       
\usepackage{amsfonts}       
\usepackage{nicefrac}       
\usepackage{microtype}      
\usepackage{graphicx}
\usepackage{amssymb}
\usepackage{amsmath}
\usepackage{enumerate}	    
\usepackage{caption} 
\usepackage{dsfont}
\usepackage{float}
\let\oldemptyset\emptyset
\newcommand{\cN}{\mathcal{N}}
\newcommand{\cL}{\mathcal{L}}

\newcommand{\cC}{\mathcal{C}}
\newcommand{\cond}{\,|\,}
\newcommand{\diff}[1]{\frac{d}{d\xi}#1}
\newcommand{\difft}[1]{\frac{d^2}{d\xi^2}#1}
\newcommand{\pdiff}[2]{\frac{\partial}{\partial #2}#1}
\newcommand{\pdifft}[2]{\frac{\partial^2}{\partial #2 ^2}#1}
\newcommand{\pgf}[1]{f_{#1}(\xi)}
\newcommand{\ptf}[1]{f_{(#1)}(\xi)}
\newcommand{\evalxi}{\Big|_{\xi=1}}
\newcommand{\gpgf}[1]{f_{#1}\big(\xi g(\zeta)\big)}
\title{Encoding Scheme for Infinite Set of Symbols:\\
       the Percolation Process on Infinite Perfect Binary Trees}

\author{
  Yousof Mardoukhi\\
  Institute of Physics and Astronomy,
  University of Potsdam,
  14476 - Potsdam, Germany \\
  \texttt{yousof.mardoukhi@uni-potsdam.de} 
}

\begin{document}
\maketitle

\begin{abstract}
It is shown here that the percolation cluster that emerges from the 
percolation process on infinite perfect binary trees, is genuinely an 
encoding scheme for an infinite set of symbols. The average codeword 
length and the entropy of such an encoding scheme are still finite as 
long as the percolation density $p$ is between $1/2\leq p < \sqrt[3]{1/4}$.
\end{abstract}

\keywords{Binary Tree\and Efficient Encoding \and Entropy 
\and Huffman Encoding Scheme \and Percolation Process \and 
Bienaym\'e-Galton-Watson Process}

\section{Introduction}
Binary trees are perennial and indispensable part of information and 
coding theory in computer science. They are used for storage file 
systems and databases~\cite{o1996, shetty2013}. For instance, the 
Btrfs (pronounced as B-Tree Filesystem) is a filesystem utilising 
the \textit{Balanced Tree} structure to store large data~\cite{rodeh2013}. 
In databases, indexing in many widely used database software such as 
PostgreSQL and Oracle SQL use B-tree indexing (see their respective 
documentations and also refer to~\cite{graefe2011} for recent methods 
and technologies). In a different area -data compression- the foundation 
of much popular data compression software such as \texttt{.arc}, \texttt{.pk(X)zip}, 
\texttt{gzip}, \texttt{.bzip} and the famous \texttt{.zip} is based on 
the \textit{Huffman encoding} scheme~\cite{huffman1952}.

In real-life cases, the source of the information is represented by a 
finite set of symbols (also called alphabets). In these situations, 
the Huffman encoding scheme ensures via an encoding procedure that the 
length of the \textit{codewords} associated to each symbol yields an 
\textit{average codeword length} which is the smallest given the frequency 
of the appearance of each symbol in the source~\cite{huffman1952}. 
Yet, one can get audacious and think of a situation where the set of 
symbols is countably infinite. Although, this would not happen in day 
to day problems, one can ask whether there exists a physical process 
that genuinely encodes the information in a manner that yields a finite 
average codeword length. For instance, one can consider a physical 
system with a set of discrete energy levels that are countably infinite. 
If one attempts to label these discrete levels with a set of codewords, 
what are the criteria such that these energy levels are expressed in 
these codewords and still on average the codeword length is finite. 

In this article, it is demonstrated that the \textit{Bienaym\'e-Galton-Watson} 
process with maximum two offspring \cite{watson1875, grimmett1989}, 
is a process that itself generates the symbols and encodes them 
simultaneously in such a manner that the state of the system described 
by these symbols can be expressed with an infinite set of codewords 
that yields a finite average codeword length. The Bienaym\'e-Galton-Watson 
process with maximum two offspring has an equivalent representation, 
namely the percolation process on infinite perfect binary trees. 
Applications of the \textit{percolation process} are far-reaching in 
many areas, ranging from soft materials such as polymers~\cite{bauhofer2009, 
kesten1987} to brittle and rigid materials such as rock~\cite{mardoukhi2017}. 
It has been used to model the financial markets~\cite{stauffer1998, yu2012} 
or division of labour markets~\cite{richardson2011} or to understand 
the dynamics of a protein in lipid membranes~\cite{almeida1992}. Yet 
with many of these applications, little has been said on the application 
of the percolation process from the perspective of information theory 
and coding theory~\cite{kobayashi2006}. It is demonstrated here that 
the \textit{percolation cluster} that emerges at the \textit{critical 
percolation density} $p_c$ has undeniable similarity to the well-known 
Huffman encoding scheme. This establishes a direct connection between 
the Bienaym\'e-Galton-Watson process, the percolation process and the 
Huffman encoding scheme.

The structure of this article is such that it first gives a gentle 
introduction to the Bienaym\'e-Galton-Watson process. Then it is shown 
that it is equivalent to the percolation process on perfect binary trees 
if the number of the offspring is limited to maximum of two descendants. 
The section afterwards is dedicated to the derivation of the generalised 
probability generating function for the Bienaym\'e-Galton-Watson process. 
With the generalised probability function in hand, one then can investigate 
the number of nodes and leaves of the percolation cluster at different 
generations. The last section concludes the results and provides the reader 
with comparisons between the analytical and the simulation results.

\section{Bienaym\'e-Galton-Watson Process and Percolation Process on Perfect Binary Trees}

The Bienaym\'e-Galton-Watson process is a branching process 
$\{\cN_n\}_{n\in\mathbb{N}}$ where $n\in\mathbb{N}$ is called 
the generation and $\cN_n$ the population of the $n^{th}$ 
generation. Each member of a generation has the possibility to have 
offspring in the next generation according to a specific probability 
distribution. The central question here is to know the number of 
the population at the $(n+1)^{th}$ generation given the number of 
the population at the $n^{th}$ generation.

In the most simplest form, it is assumed that the number of offspring 
each member of a generation has is independent of the other members of 
that generation or any other generations before. If the number of offspring 
the $i^{th}$ member of the $n^{th}$ generation has is denoted by 
$X_n^i\in\mathbb{N}$, then the assumption made earlier implies that 
$X_n^i$'s are \emph{i.i.d} random variables according to a given fixed 
probability distribution. Therefore, 
\begin{align}
	\cN_{n+1} = \sum_{i=1}^{\cN_n} X_n^i.
\end{align}
It follows immediately that the transition probability function for the 
process is as follows~\cite{asmussen1983}
\begin{align}
	P\big(\cN_{n}=i\,|\,\cN_{n-1} = j\big) = 
	\begin{cases}
		p_i^{*j},& \text{if}\quad i\geq1,\,j\geq0\\
		\delta_{0i},& \text{if}\quad i=0,\,j\geq0 
	\end{cases},
	\label{eq:ptf}
\end{align}
where $p_i^{*j}$ is understood as the $j$-fold probability convolution. 
Note that this probability transition function implies that the process is 
a Markov chain. 

\begin{figure}[!h]
	\centering
	\includegraphics[width=.3\textwidth]{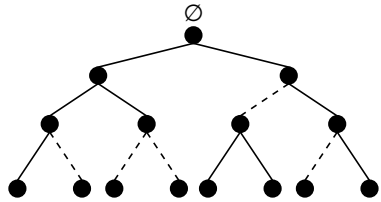}
	\caption{A perfect binary tree with $n=3$. The percolation process 
	on this tree left behind a set of open edges (solid lines) along with 
	closed edges (dashed lines) with probability $p$ and $q$ respectively.
	}
	\label{fig:perconbint}
\end{figure}

There is a kin relation between the Bienaym\'e-Galton-Watson process and 
the percolation process on trees. Since the ultimate goal of this article 
is about information encoding, henceforth only the binary trees are assumed
here. Specifically, consider infinite perfect binary trees with the root 
$\emptyset$ at the top. The percolation process on such trees is the process 
of assigning to each edge independent of any other edges the state of being 
open with probability $p$and the state of being closed with probability 
$q=1-p$. The probability $p$ is also known as the percolation density. The 
nodes in the immediate vicinity of the root belong to the first generation 
and are the offspring of the root. The nodes that are in the immediate 
vicinity of the nodes in the first generation belong to the second generation 
and so on. A realisation of such a process on a perfect binary tree up to the 
third generation is shown in Fig.~\ref{fig:perconbint}.
An open path between a node in the $n^{th}$ generation and a node in the 
$m^{th}$ generation is a sequence of open edges that starts at the former 
node and ends at the latter. Consequently, two nodes are said to be connected 
if there is an open path between them. A set of nodes that are connected 
form an \emph{open cluster}. The size of an open cluster is simply the 
number of nodes it has. A node is also said to be a leaf if it does not 
have any offspring.

When $p=1$, a single perfect tree emerges and all the nodes belong 
to one unique open cluster. On the other hand, when $q=1$, nodes are 
trivial trees. In this case, one has a forest in which the nodes are the 
trivial trees in this forest. In Bienaym\'e-Galton-Watson process if the 
probabilities of having offspring more than two are zero, then every open 
cluster of the percolation process is a realisation of a 
Bienaym\'e-Galton-Watson process that commences at a node which is not the 
offspring of any other node.

The interesting fact here though is the emergence of an infinite open 
cluster that includes the root when $p$ reaches a critical value called $p_c$ 
when for the first time an open cluster called the \textit{percolation 
cluster} emerges that is infinite in size when $n\rightarrow\infty$. This 
corresponds to the critical Bienaym\'e-Galton-Watson process when it thrives 
ad infinitum. It is possible to investigate the distribution of the nodes and 
leaves at $p_c$ (also for any other $p$) using the known tool of probability 
generating functions~\cite{harris1963, drmota1994} discussed in the following 
subsection.

\subsection{Probability generating function of the Bienaym\'e-Galton-Watson
process}
\label{sec:pgf}

The probability generating function for the Bienaym\'e-Galton-Watson 
process over the states $\mathbb{N}=\{0, 1, 2, \cdots\}$ and the associated
probabilities $\{p_k\}_{k\in\mathbb{N}}$ is defined by 
\begin{align}
	f(\xi) = \sum_{k=0}^\infty p_k \xi^k,\quad |\xi|\leq 1,
	\label{eq:pgf}
\end{align}
where the dummy index $k$ counts the number of offspring and $p_k$ is the 
probability of having $k$ offspring. Note that in the case of binary trees,
$p_k = 0$ for all $k > 2$. Hence, the upper bound of the sum is identically
2. Nonetheless, the results derived henceforth in this section are not limited 
to this constraint. The iterates of the probability generating function 
Eq.~\eqref{eq:pgf} are given by
\begin{align}
	\nonumber
	f_0(\xi)=\xi, \quad 
	f_1(\xi) = f(\xi), \quad 
	f_n(\xi) = f[f_{n-1}(\xi)].
\end{align}
It is not difficult to establish the following using Eq.~\eqref{eq:ptf}
\begin{subequations}
\begin{gather}
	\sum_kP(\cN_n = k\cond\cN_{n-1} = 1)\xi^k = 
	\sum_k p_k\xi^k = f(\xi),\\
	\sum_kP(\cN_n = k\cond\cN_{n-1} = i)\xi^k = 
	[f(\xi)]^i.
\end{gather}
\end{subequations}
The most central identity concerning the probability generating function   
defined in Eq.~\eqref{eq:pgf} and the probability transition function given
by Eq.~\eqref{eq:ptf} is the one that establishes a relation between the 
$n^{th}$ iterate of the probability generating function $f_n(\xi)$ and 
the probability generating function associated with the $n^{th}$ step of the 
probability transition function, denoted by $f_{(n)}(\xi)$. Due to 
Chapman-Kolmogorov equation~\footnote{also known as Smoluchowski equation} 
for Markov chains observe that
\begin{align}
	\nonumber
	f_{(n)}(\xi) & = \sum_k P(\cN_n=k\cond\cN_{0}=1)\xi^k =
	\sum_k\sum_s P(\cN_{n-1}=s\cond\cN_0=1)
		     P(\cN_n=k\cond\cN_{n-1}=s)\xi^k\\
	\nonumber
	& = \sum_s P(\cN_{n-1}=s\cond\cN_0=1)\sum_k 
	           P(\cN_n=k\cond\cN_{n-1}=s)\xi^k\\
	& = \sum_s P(\cN_{n-1}=s\cond\cN_0=1)[f(\xi)]^s 
	  = f_{(n-1)}[f(\xi)] 
	  = \underbrace{f[\cdots[f(\xi)]]}_{n\text{times}} = \pgf{n}
	\label{eq:equividen}
\end{align}
The identity above is crucial to deduce moments and properties of the 
Bianaym\'e-Galton-Watson process. In this regard, first note that the 
mean and the variance of the population in the first generation can be 
calculated via the probability generating function $f(\xi)$.  
\begin{subequations}
\begin{align}
	&\diff{f(\xi)}\evalxi = \sum_{k=0}^2 k p_k = 
		\mathbb{E}[\cN_1]:=\mu,
	\label{eq:diffpgf}\\
	&\difft{f(\xi)}\evalxi = 
		\sum_{k=0}^2 k(k-1)p_k = \mathrm{Var}[\cN_1]-\mu + \mu^2.
	\label{eq:difftpgf}
\end{align}
\end{subequations}
Moreover
\begin{align}
	\nonumber
	\mathbb{E}[\cN_n] & = \diff{\ptf{n}}\evalxi = 
	\diff{\pgf{n}}\evalxi = \diff{f[\pgf{n-1}]}\evalxi \\
	& = f^\prime[\pgf{n-1}]f^\prime_{n-1}(\xi)\evalxi = 
	f^\prime(1)f^\prime_{n-1}(1) = [f^\prime(1)]^n = \mu^n,
	\label{eq:Nmean}
\end{align}
where the prime in $f^\prime(.)$ stands on the derivative with respect to 
the argument. In the last step, one recursively invokes the same identity 
for $f^\prime_{n-1}(\xi)$ and arrives at $\mu^n$. Follow the same procedure
and deduce that
\begin{align}
	\nonumber
	\mathrm{Var}[\cN_n]-\mu^n+\mu^{2n} 
	&= \difft{\ptf{n}}\evalxi = \difft{\pgf{n}}\evalxi
	= \difft{f[\pgf{n-1}]} \\
	\nonumber
	&= f''[f_{n-1}(\xi)](f'_{n-1}(\xi))^2\evalxi 
	+ f'[f_{n-1}(\xi)]f''_{n-1}(\xi)\evalxi\\
	&= f''(1)\mu^{2n-2} + \mu f''_{n-1}(1) 
	= f''(1)\mu^{2n-2}\sum_{i=0}^{n-1}\mu^{-i}.
	\label{eq:Nvar1}
\end{align}
The last line in the equation above is due to applying recursively 
the identity $f''_n(1) = f''(1)\mu^{2n-2} + \mu f''_{n-1}(1)$. Depending on
the value of $\mu$ and by substituting $f''(1)$ with 
Eq.~\eqref{eq:difftpgf} the variance obeys the following expressions
\begin{align}
	\mathrm{Var}[\cN_n] = 
	\begin{cases}
	   \mathrm{Var}[\cN_1]\mu^n\left(\frac{\mu^n-1}{\mu^2-\mu}\right),&
		\text{if}\quad \mu\neq 1\\
	   n\mathrm{Var}[\cN_1],& \text{if}\quad \mu=1
	\end{cases}.
	\label{eq:Nvar2}
\end{align}
So far, the discussion was focused on the statistical properties of $\cN_n$
. In the picture of the percolation process on infinite perfect binary 
trees $\cN_n$'s correspond to the number of nodes at different generations.
Yet, the number of those nodes without any offspring (called leaves) is of 
utmost importance in the upcoming section. The number of leaves at 
generation $n$ is a random variable $\mathcal{L}_n$ and it is formally 
given by the following sum.
\begin{align}
	\nonumber
	\cL_n = \sum_{i=1}^{\cN_n} \mathds{1}_L(i),
\end{align}
where $\mathds{1}_L(i)$ is an indicator function such that 
$\mathds{1}_L(i)=1$ if $i$ belongs to the set of leaves 
$\overline{L}\subset\{\cN_n\}$ and is zero otherwise. The notation $\{\cN_n\}$ 
indicates the set of the population at the $n^{th}$ generation. The probability 
transition function for $\cL_n$ is read as
\begin{align}
	\mathrm{Pr}(\cL_n = i\cond, \cN_n = j \wedge \cN_{n-1} = k) = 
	P(\cL_n = i \cond \cN_n=j)P(\cN_n = j \cond \cN_{n-1} = k).
\end{align}
This implies that the probability of observing $i$ leaves at generation $n$
is the joint probability of having $j$ nodes at the same generation given 
that the number of nodes in the previous generation is $k$, and from those 
$j$ nodes $i$ of them are leaves i.e. 
\begin{align*}
	&P(\cL_n = i \cond \cN_n = j \wedge \cN_{n_1} = 1) = \\ 
	&P(\cL_n = i \cond \cN_n = j)P(\cN_n = j \cond \cN_{n-1} = 1)
	= \binom{j}{i}u_1^i u_0^{j-i} p_j.
\end{align*}
Here $u_1$ is the probability of being a leaf whilst $u_0$ is the 
probability of not being a leaf. The notation $\binom{j}{i}$ stands on 
choosing $i$ elements from $j$ elements. Multiply both sides by 
$\xi^j\zeta^i$ and sum over $i$ and $j$ and arrive at the following.
\begin{align}
	\nonumber
	&\sum_{i,j} P(\cL_n = i \cond \cN_n = j)
	P(\cN_n = j \cond \cN_{n-1} = 1)= \\
	&\sum_{j}p_j\xi^j\sum_i\binom{j}{i}u_1^iu_0^{j-i}\zeta^i 
	= \sum_j p_j [\xi(u_1\zeta + u_0)]^j = f\big(\xi g(\zeta)\big),
	\label{eq:gpgf}
\end{align}
where $g(\zeta) = (u_1\zeta + u_0)$ is the probability generating 
function of the state of a single node being a leaf or not and 
$f\big(\xi g(\zeta\big)$ is the \emph{generalised probability generating 
function} that includes the information of both the number of nodes and 
leaves at a given generation $n$. 

The generalised probability generating function satisfies the same 
properties that $f(\xi)$ exhibits. For instance
\begin{align}
	\nonumber
	\gpgf{(n)} &= \sum_{i,j}P(\cL_n=i\cond\cN_n=j\wedge\cN_0=1)
		    \xi^j\zeta^i\\
	\nonumber
	&= \sum_{i,j}P(\cL_n=i\cond\cN_n=j)P(\cN_n=j\cond\cN_0=1)
		     \xi^j\zeta^i\\
	\nonumber
	&= \sum_{i,j}P(\cL_n=i\cond\cN_n=j)\sum_k
	             P(\cN_n=i\cond\cN_{n-1}=k)P(\cN_{n-1}=k\cond\cN_0=1)
		     \xi^j\zeta^i\\
	&= \sum_k P(\cN_{n-1}=k\cond\cN_0=1)
	   \left[f\big(\xi g(\zeta)\big)\right]^k 
	   = f_{(n-1)}\left[f\big(\xi g(\zeta)\big)\right] = \gpgf{n},
\end{align}
where the last term is followed by induction. The average number of leaves 
at a given generation $n$ is achieved by taking the first order partial 
derivative of $\gpgf{(n)}$ with respect to $\zeta$
\begin{align}
	\nonumber
	\mathbb{E}[\cL_n] &= \pdiff{\gpgf{(n)}}{\zeta}\Big|_{\xi=\zeta=1}
	= \pdiff{\gpgf{n}}{\zeta}\Big|_{\xi=\zeta=1}
	= \pdiff{f[\gpgf{n-1}]}{\zeta}\Big|_{\xi=\zeta=1}\\
	\nonumber
	& = f'\left[\gpgf{n-1}\right]f'_{n-1}\big(\xi g(\zeta)\big)\xi 
	    g'(\zeta)\Big|_{\xi=\zeta=1}\\
	&= u_1f'(1)f'_{n-1}(1) = u_1 [f'(1)]^n = u_1\mathbb{E}[\cN_n]
	 = u_1\mu^n,
	\label{eq:Lnmean}
\end{align}
where in the last line the identity that $f'_{n}(1) = [f'(1)]^n$ derived in
Eq.~\eqref{eq:Nmean} is used. In a similar fashion used in 
Eq.~\eqref{eq:Nvar1} the variance of $\cL_n$ is deduced to be
\begin{align}
	\mathrm{Var}[\cL_n]=u_1^2\mathrm{Var}[\cN_n].
	\label{eq:Lnvar}
\end{align}

\subsection{The percolation process on perfect binary trees and the 
statistical properties of $\cN_n$ and $\cL_n$}
\begin{figure}
	\centering
	\includegraphics[width=.8\textwidth]{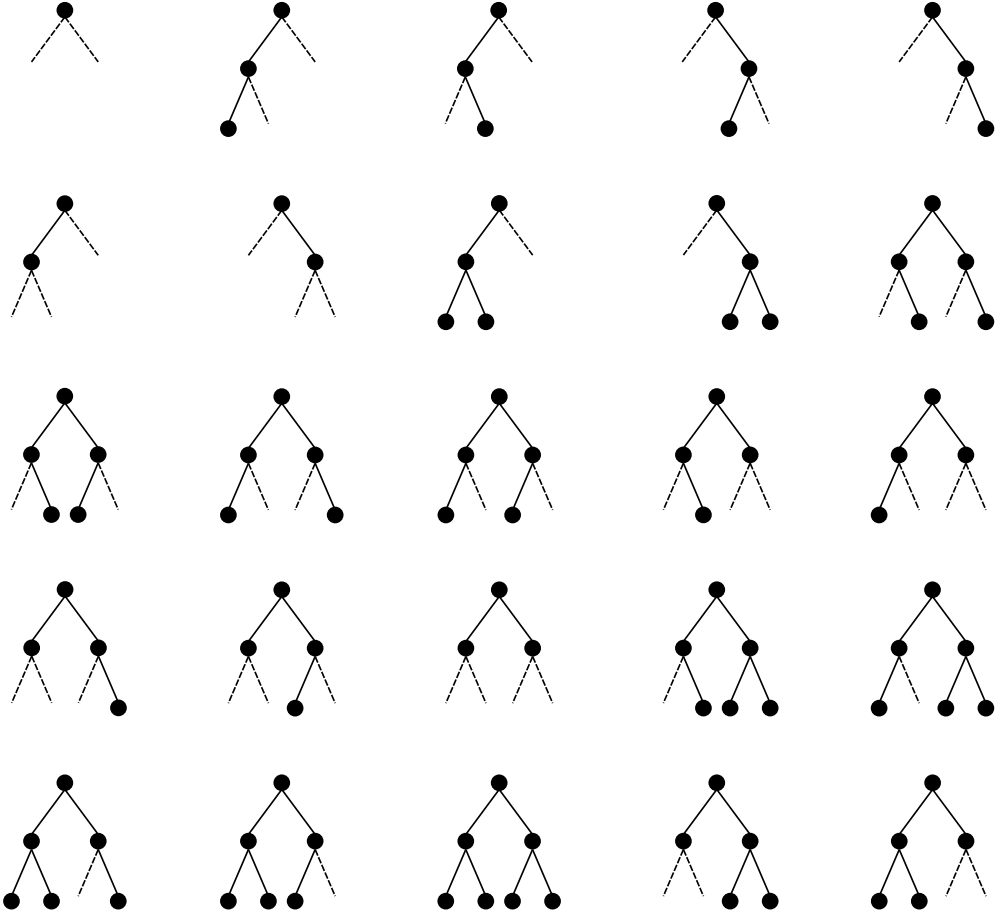}
	\caption{All possible configurations of nodes at $n=1$. 
	Solid lines represent edges that are open with probability 
	$p$ and dashed lines represent edges that are blocked with 
	probability $q$. These are all the configurations that the 
	Eq.~\eqref{eq:gpgfg} generates. }
	\label{fig:bin_config}
\end{figure}

As previously mentioned, percolation process is equivalent to the Bienaym\'e
-Galton-Watson process discussed earlier. If one limits the number of 
offspring of each node only to the set $\{0, 1, 2\}$, then one deals with 
the percolation process on perfect binary trees. The fact that a node does
not have any offspring corresponds to the situation where the edges coming 
out of that very node are both closed which corresponds to the probability 
$q^2$. Likewise, the possibility of having only one offspring corresponds 
to the situation where one edge is closed and the other one is open which 
implies the probability $2pq$ and having exactly two offspring indicates 
that both edges are open and the probability associated to this event is 
$p^2$. Hence one identifies the probabilities $p_0, p_1$ and $p_2$ with 
$q^2, 2pq$ and $p^2$ respectively. Thus, the probability generating function 
$f(\xi)$ given by Eq.~\eqref{eq:pgf} in closed form is written as
\begin{align}
	f(\xi) = \sum_{i=0}^2 p_i\xi^i 
	= q^2 + 2pq\xi + p^2\xi^2 = (p\xi+q)^2.
	\label{eq:npgf}
\end{align}
Moreover, a node is a leaf if the edges coming out of it are both closed, 
corresponding to the probability $q^2$. Whilst, it is not a leaf if it has 
at least one offspring which corresponds to the probability $2pq + p^2$. 
Therefore, the probability generating function of leaves, namely $g(\zeta)$
is given by
\begin{align}
	g(\zeta) = \sum_{i=0}^1 u_i\zeta^i = q^2\zeta + 2pq + p^2.
	\label{eq:lpgf}
\end{align}
Using Eq.~\eqref{eq:npgf} and Eq.~\eqref{eq:lpgf}, the generalised probability 
generating function $f\big(\xi g(\zeta)\big)$ is given by
\begin{align}
	f\big(\xi g(\zeta)\big) = \sum_{i=0}^2 p_i\big(\xi g(\zeta)\big)^i
	= \big(p\xi g(\zeta)+q\big)^2.
	\label{eq:gpgfg}
\end{align}
All the possible configurations of the nodes that this generalised 
probability function generates are listed in Fig.~\ref{fig:bin_config}.
It is straightforward to calculate the mean and variance of $\cN_1$ thanks to 
Eq.~\eqref{eq:diffpgf} and Eq.~\eqref{eq:difftpgf},
\begin{subequations}
\begin{align}
	&\mathbb{E}[\cN_1]  = \mu
	= \pdiff{f \big(\xi g(\zeta)\big) }{\xi}\Big|_{\xi=\zeta=1}
	= 2pq + 2p^2 = 2p\\ 
	&\mathrm{Var}[\cN_1] 
	= \pdifft{ f\big(\xi g(\zeta) \big) }{\xi}\Big|_{\xi=\zeta=1} 
	- \mu^2 + \mu = 2p^2 - (2p)^2 + 2p = 2p(1-p) = 2pq.
\end{align}
\end{subequations}
Then due to Eq.~\eqref{eq:Nmean} and Eq.~\eqref{eq:Nvar2} the followings 
are yielded
\begin{subequations}
\begin{align}
	&\mathbb{E}[\cN_n] = (\mathbb{E}[\cN_1])^n = (2p)^n,\\
	&\mathrm{Var}[\cN_n] = 
	\begin{cases}
	   \mathrm{Var}[\cN_1]\mu^n\left(\frac{\mu^n-1}{\mu^2-\mu}\right) 
	   = q(2p)^n\left[\frac{(2p)^n-1}{2p-1}\right],& \text{if}\quad p\neq1/2\\
	   n\mathrm{Var}[\cN_1] = 2npq = \frac{n}{2},& \text{if}\quad p=1/2
	\end{cases}.
\end{align}
\end{subequations}
By virtue of equations~\eqref{eq:Lnmean} and~\eqref{eq:Lnvar} and the 
results of the equations above, the mean and the variance of $\cL_n$ are 
given by
\begin{subequations}
\begin{align}
	&\mathbb{E}[\cL_n] = q^2\mathbb{E}[\cN_n] = q^2(2p)^n,\\
	&\mathrm{Var}[\cL_n] = q^2\mathrm{Var}[\cN_n] =
	\begin{cases}
	  q^3(2p)^n\left[\frac{(2p)^n-1}{2p-1}\right],
		&\text{if}\quad	p\neq 1/2\\
	  2npq^3 = \frac{n}{8},&\text{if}\quad p=1/2
	\end{cases}.
\end{align}
\end{subequations}
It is worthwhile to discuss for which $p$ the percolation cluster emerges. 
It is clear that the emergence of the percolation cluster is associated to 
the critical Bienaym\'e-Galton-Watson process. Equation~\eqref{eq:Nmean} 
implies that when $\mu < 0$, the expectation value of $\cN_n$ tends to 
zero as $n\rightarrow\infty$ and for $\mu >0$, $\cN_n$ diverges conversely.
Yet, for $\mu=1$, for all the generations, the expectation value remains unity.
This case implies that for the percolation process $p=1/2$, since 
$\mu=2p=1$. Hence, for $p<1/2$ all the open clusters for a percolation 
process on a perfect binary tree are finite in size and when $p>1/2$ a 
unique percolation cluster exists. The value $p=1/2$ is also deducible by 
solving the equation $\xi = f(\xi)$.
\begin{align*}
	(p\xi + q)^2 = \;&\xi \rightarrow \xi 
	= \frac{2p^2 + 1 - 2p - |1 - 2p|}{2p^2}\\
	&\xi = 
	\begin{cases}
		1\quad &\mathrm{if}\quad p \leq 1/2, \\
		(q/p)^2\quad &\mathrm{if}\quad p > 1/2
	\end{cases}.
\end{align*}
When $p < 1/2$, the probability that the percolation cluster emerges is zero 
whilst when $p > 1/2$, the chance that it does not appear is given by 
$(q/p)^2$~\cite{grimmett1989}. 

\section{Maximal encoded information}
The percolation clusters on perfect binary trees discussed in the previous
section are unquestionably related to the Huffman encoding scheme which is 
used to efficiently encode symbols into strings of 0's and 1's 
algorithmically. Binary encoding is the process of assigning a binary 
string to a set of symbols. For instance, the English alphabets \textit{A} 
to \textit{Z} can be represented by strings of 0's and 1's. An example of 
such an encoding procedure is shown in Tab.~\ref{tab:fixed-length}. 
\begin{table}
	\begin{center}
	\begin{tabular}{||c | c||}
		\hline
		Symbol & Encoded \\ [0.5ex]
		\hline\hline
		A & 00000 \\
		\hline
		B & 00001 \\
		\hline
		C & 00010 \\
		\hline
		D & 00011 \\
		\hline
		\vdots & \vdots \\
		\hline
		Z & 11010 \\ [0.1ex]
		\hline
	\end{tabular}
	\end{center}
	\caption{One possible fixed-length encoding scheme for 
		the English alphabets. }
	\label{tab:fixed-length}
\end{table}
This encoding is called \textit{fixed-length} encoding scheme~\cite{roman1996}. 
Though convenient, it is not the best and efficient encoding procedure as 
there are many unnecessary 0's and 1's that are used to represent the symbols. 
Another supplementary piece of information that indeed helps to make the encoding 
more efficient is the fact that the English alphabets have different frequencies 
of appearance in text sources. For instance, the letter $E$ with 12.02\% has 
the highest frequency of appearance and the letter $Z$ with 0.07\% has the 
least. Thus, it is logical to represents the letter $E$ with fewer bits of 0's 
and 1's, e.g. with only one single digit 0.

One defines an information source as an ordered pair 
$\mathcal{S} = (S, P)$ where $S$ is the set of symbols e.g. 
$S=\{s_1, s_2, \cdots, s_n\}$ and $P$ is a probability measure 
$P=\{p_{s_1}, p_{s_2}, \cdots, p_{s_n}\}$, where $p_{s_i}$ is the 
probability of appearance of the symbol $s_i$ in the source. An encoding scheme 
is then an ordered pair $\mathcal{E} = (C, f_c)$ where $C$ is the set of codes e.g. 
$\{00, 01, 000, 101, \cdots\}$ and $f_c$ is the \textit{encoding function} 
$f_c: S\rightarrow C$ which maps a symbol from the set $S$ to a \textit{codeword} in 
$C$. The average codeword length for an information source $\mathcal{S}$ and the 
encoding scheme $\mathcal{E}$ is defined as 
\begin{align}
	L = \sum_{i=1}^{n} p_{s_i} \mathrm{Len}\big(f_c(s_i)\big),
	\label{eq:avgcodel}
\end{align}
where $\mathrm{Len}\big(f_c(s_i)\big)$ is the length of the codeword 
associated with the symbol $s_i$ via the map $f_c(.)$.

The average codeword length is a quantity which its magnitude measures how 
efficient an encoding scheme is. Obviously, the smaller the $L$ is, the 
more efficient the encoding is as lesser bits of 0's and 1's are required 
to represent the message formed by the set of the symbols. Note that $L$ 
cannot be arbitrarily small as it is required that the encoding scheme 
$\mathcal{E}$ to be \textit{uniquely decipherable} or even more 
desirable, to be \textit{instantaneous}~\cite{roman1996}. 
Huffman devised an algorithm which is now known after his name that yields 
the most efficient encoding scheme ensuring that the scheme is 
instantaneous~\cite{huffman1952}. The lower bound for $L$ was proved to be 
the amount of the information in $\mathcal{S}$ which is given by the 
Shannon's entropy defined as~\cite{roman1996}
\begin{align}
	\mathcal{H} = -\sum_{i=1}^{n} p_{s_i}\log (p_{s_i}) \leq L.
	\label{eq:shanntropy}
\end{align}

\subsection{Percolation cluster as an efficient encoding scheme}
\begin{figure}[!ht]
	\centering
	\includegraphics[width=.3\textwidth]{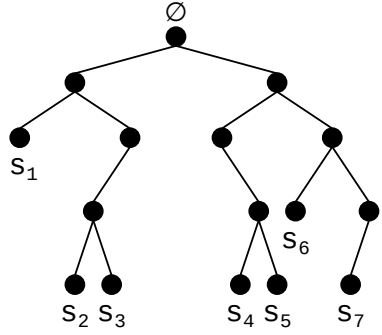}
	\caption{An open cluster with $n=4$ generated by some percolation 
	process. Only the open edges are shown. The leaves represents the 
	symbols. Every edge to the left can be represented by the bit 0 and 
	to the right with 1. }
	\label{fig:perc_bin_encod}
\end{figure}

The percolation cluster, though not an encoding scheme per se, can be 
regarded as encoding scheme which assign to a set of symbols (leaves) a set
of probabilities of appearance. Thus, one can say that the percolation 
cluster contains the information source $\mathcal{S}$ and encodes it 
simultaneously according to the following procedure
\begin{enumerate}[(i)]
	\item every leaf represents a symbol which has a probability 
associated with identified by the Bernoulli probability measure 
$\prod_{i}^{n} p$  where $n$ is the generation at which the leaf resides.

	\item the leaf is encoded as a binary string of length $n$. The 
string is generated by mapping every turn to the left to the digit 0 and 
every turn to the right to the digit 1 when traversing the open path from 
the root $\oldemptyset$ to the leaf similar to the procedure in Huffman 
encoding.
\end{enumerate}
To elaborate this further, consider an imaginary open cluster generated 
by an arbitrary percolation process depicted in Fig.~\ref{fig:perc_bin_encod}.
In this figure the open cluster has 7 leaves at different 
generations. Every leaf in that very process can be associated to a 
single symbol. One can further intuitively assign to each symbol a 
unique codeword by traversing the open path which starts at the root 
and ends at the symbol. Every turn to the left represents the bit 0 and 
to the right the bit 1. Hence, the open cluster automatically 
provides an encoding scheme $\mathcal{E}$ which maps the symbols 
$S = \{s_1, s_2, \cdots, s_7\}$ according to the one-to-one correspondence 
shown in Tab.~\ref{tab:perc_coding}.
\begin{table}
	\begin{center}
	\begin{tabular}{||c | c||}
		\hline
		Symbols & Codewords \\ [0.5ex]
		\hline\hline
		$s_1$ &  00\\
		\hline
		$s_2$ & 0100 \\
		\hline
		$s_3$ & 0101 \\
		\hline
		$s_4$ & 1010 \\
		\hline
		$s_5$ & 1011\\
		\hline
		$s_6$ & 110 \\ 
		\hline
		$s_7$ & 1110 \\ [0.1ex]
		\hline
	\end{tabular}
	\end{center}
	\caption{Instantaneous encoding yielded by the open cluster in 
	Fig.~\ref{fig:perc_bin_encod} yielded by a percolation process. }
	\label{tab:perc_coding}
\end{table}
This encoding is \textit{instantaneous}. This is deduced by the fact that 
every instantaneous encoding has binary tree representation (refer to
~\cite{reina2014} and the Kraft's and McMillan's theorem for instance 
in~\cite{roman1996}).

Since the appearance of each leaf is independent of the other leaves, a 
natural Bernoulli probability measure can be associated to each leaf. Thus,
for instance, the symbol $s_1$ has the probability of appearance 
proportional to $\prod_{i=0}^{2}p$ and the symbol 
$s_7$, $\prod_{i=0}^{4} p$. Hence, the probability of the codewords is 
spontaneously yielded by the open cluster with the extra caution that
it has to be normalised.  

Therefore, every open cluster generated via a percolation process is 
inherently an encoding scheme in which the information source $\mathcal{S}$
is the ordered pair of the set of symbols $S$ identified by the leaves and 
the probability distribution $P$ which is identified by the Bernoulli 
measure associated to the length of the open path starting from the root and 
ending at the leaves. The codeword for each symbol is understood as the sequence 
of 0's and 1's where the digit 0 represents the open edges to the left and the 
digit 1 represents the open edges to the right. Subsequently, the encoding 
function $f_c$ is identified by the correspondence between the symbols and 
the sequence of 0's and 1's which is yielded by traversing the open path from 
the root to the leaf representing the symbol.

\subsection{Encoding a countably infinite set of symbols}
When the set of symbols $S$ is finite, given that $P$ is a probability 
measure and the length of the codewords in $C$ is finite, the Shannon's 
entropy and the average codeword length are both well-defined and finite. 
Yet, a legit question would be whether one can efficiently encode a 
set of symbols that is countably infinite. As discussed earlier, a finite 
cluster formed when $p<1/2$ can be thought of an encoding scheme on a 
finite set of symbols. Au contraire, when $p>1/2$ the percolation cluster 
emerges and the number of leaves goes to infinity and hence one can 
consider the percolation cluster as an encoding scheme on an infinite set 
of symbols. For a given realisation $\cC$ of a percolation cluster, the 
Shannon's entropy Eq.~\eqref{eq:shanntropy} can be rewritten as follows
\begin{align}
	\mathcal{H}(\cC) 
	= -\sum_{i=1}^{\infty} 
	  p\left(s_i\cond\cC\right)\log_2\big(p(s_i\cond\cC)\big) 
	= -\sum_{n=0}^{\infty} \cL_n(\cC) p^n \log_2(p^n),
\end{align}
where now instead of having the dummy index of the sum counting the index 
of the symbols, it goes through different generations of the tree. 
The probabilities are factorised by the number of leaves at every given 
generation $n$. The problem here is that the set $P$ does not sum to unity
and hence not a probability measure. Hence, it is required to normalise the
probabilities i.e.
\begin{align}
	\mathcal{H}(\cC) =
	-\sum_{n=0}^{\infty} \frac{\cL_n(\cC) p^n}{\Lambda(\cC)} 
	\log_2\left(\frac{p^n}{\Lambda(\cC)}\right),
	\label{eq:entperc}
\end{align}
where $\Lambda(\cC) = \sum_{n=0}^\infty \cL_n(\cC) p^n$ is the 
normalisation factor for a given configuration $\cC$. Clearly, 
$\Lambda(\cC)$ is a random variable that depends on the configuration $\cC$
. Therefore, taking the average of the entropy $H(\cC)$ involves taking 
average over $\Lambda(\cC)$ as well. Yet, observe that
\begin{align*}
	&\mathbb{E}[\Lambda(\cC)] := \lambda
	= \sum_{n=0}^\infty \mathbb{E}[\cL_n(\cC)]p^n
	= q^2\sum_{n=0}^\infty (2p^2)^n = \frac{q^2}{1-2p^2},
	\quad \text{when}\quad 1/2 \leq p < \sqrt{1/2}\\
	&\mathrm{Var}[\Lambda(\cC)] 
	= \sum_{n=0}^\infty \mathrm{Var}[\cL_n]p^n 
	= \frac{q^3}{2p-1} \sum_{n=0}^\infty (2p^2)^n\left[(2p)^n-1\right]\\
	&\qquad= \frac{q^3}{2p-1} 
	\left[\frac{1}{1-4p^3} - \frac{1}{1-2p^2}\right] 
	= \frac{2p^2q^3}{(1-4p^3)(1-2p^2)}
	\quad \text{when}\quad 1/2 \leq p < \sqrt[3]{1/4},\\
	&\mathrm{Var}[\Lambda(\cC)]
	= \sum_{n=0}^\infty \mathrm{Var}[\cL_n]p^n 
	= 2pq^3\sum_{n=0}^\infty n p^n  
	= 2qp^2 = 1/4\quad\text{when}\quad p=1/2 
\end{align*}
Therefore, as long as $1/2 \leq p < \sqrt[3]{1/4}$ the mean and the 
variance of $\Lambda$ are finite and well-defined. Thus, as 
$n\rightarrow\infty$ it is sound to substitute $\Lambda(\cC)$ with its mean
$\lambda$. This allows one to rewrite Eq.~\eqref{eq:entperc} as follows
\begin{align}
	\mathcal{H}(\cC) =
	-\sum_{n=0}^{\infty} \frac{\cL_n(\cC) p^n}{\lambda} 
	\log\left(\frac{p^n}{\lambda}\right).
\end{align}
Consequently,
\begin{align}
	\nonumber
	\mathbb{E}[\mathcal{H}(\cC)] &=
	-\sum_{n=0}^{\infty} \frac{\mathbb{E}[\cL_n(\cC)] p^n}{\lambda} 
	\log\left(\frac{p^n}{\lambda}\right)
	= -\frac{q^2}{\lambda} \sum_{n=0}^{\infty}(2p^2)^n\log(p^n/\lambda)
	\\ &= -\frac{q^2}{\lambda} \Big[\sum_{n=0}^{\infty}n(2p^2)^n\log(p)
	  - \sum_{n=0}^{\infty}(2p^2)^n\log(\lambda)\Big] 
	= \frac{2p^2\log(1/p)}{1-2p^2} + \log(\lambda).
	\label{eq:avg_entropy}
\end{align}
Similarly, for the average codeword length,
\begin{align}
	\mathbb{E}[L(\cC)] 
	= \sum_{n=0}^{\infty} \frac{n\mathbb{E}[\cL_n(\cC)] p^n}{\lambda}
	= \frac{q^2}{\lambda} \sum_{n=0}^{\infty}n(2p^2)^n
	= \frac{2p^2}{1-2p^2}
	\label{eq:avg_codewordlength}
\end{align}

This demonstrates that the percolation cluster formed via the percolation 
process on infinite perfect binary trees equipped with the Bernoulli 
probability measure, encodes the leaves of the percolation cluster in such 
a way that the entropy and the average codeword length remain finite, 
although the number of leaves tend to infinity when 
$1/2 \leq p < \sqrt[3]{1/4}$.

\begin{figure}
	\centering
	\begin{minipage}{.43\textwidth}
		\includegraphics[width=\textwidth]{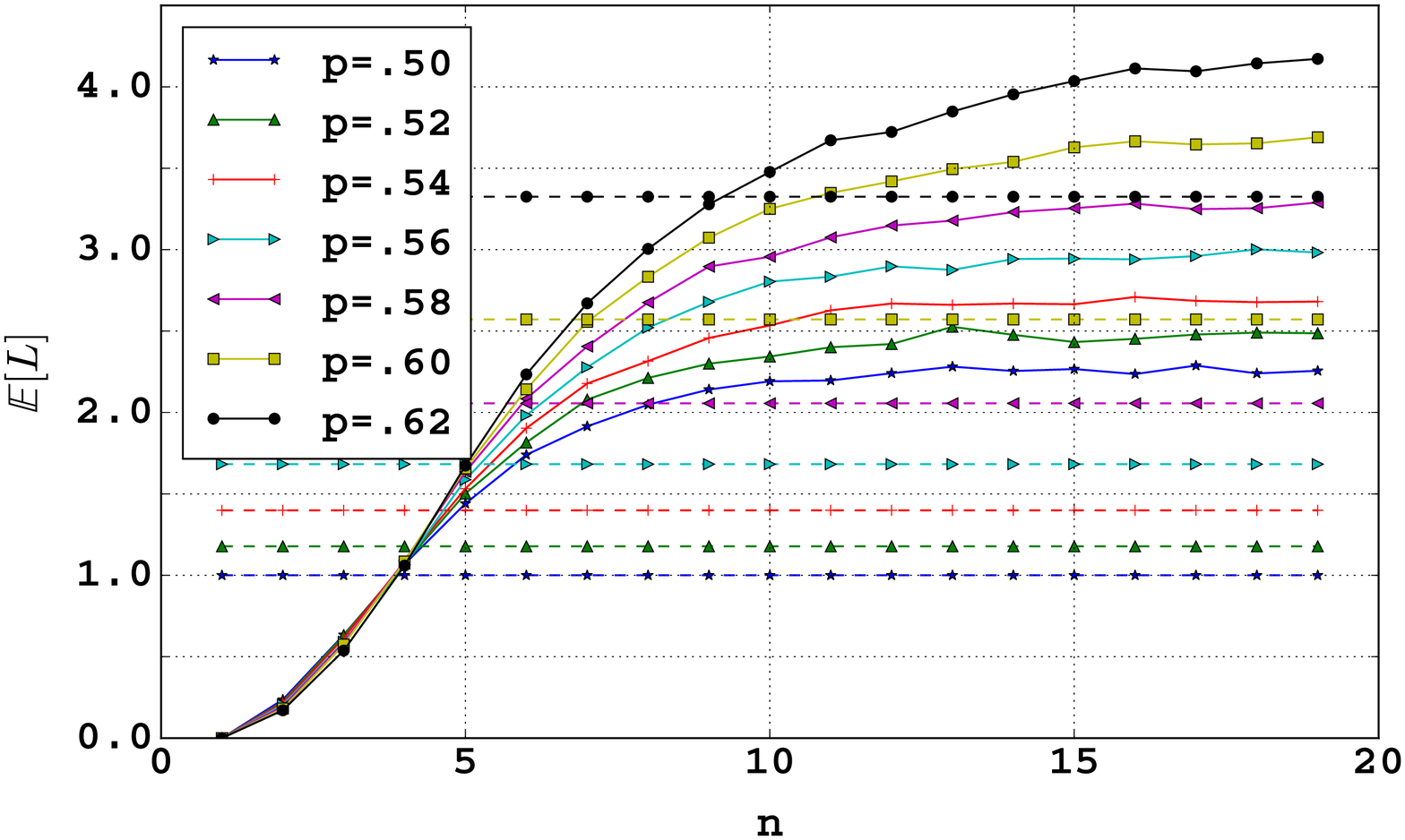}
	\end{minipage}\qquad
	\begin{minipage}{.43\textwidth}
		\includegraphics[width=\textwidth]{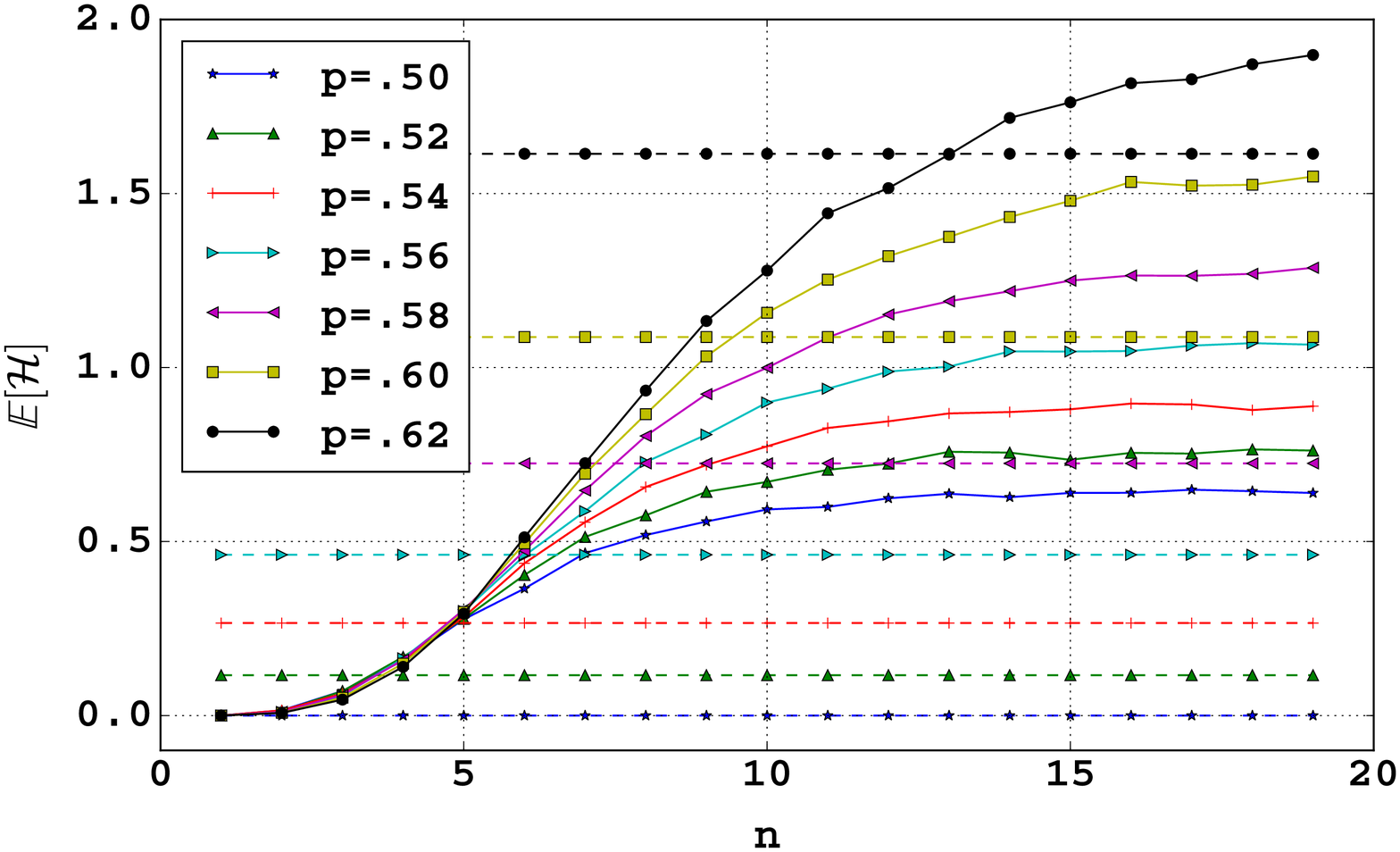}
	\end{minipage}
	\caption{(left) The average codeword length for various percolation
		 densities $p$ versus the depth of the tree $n$. It is seen
		 that $\mathbb{E}[L]$ saturates as long as 
		 $ 1/2\leq p < 1/\sqrt[3]{4}$. (right) The same as the left
		 pain but for the average entropy $\mathbb{E}[\mathcal{H}]$
		 . The solid lines represent the simulation results while 
		 the dashed lines are the saturation plateaus predicted by
		 the Eq.~\eqref{eq:avg_codewordlength} and 
		 Eq.~\eqref{eq:avg_entropy}. }
	\label{fig:avg_length_entropy}
\end{figure}

These results are also tested against the simulations. The simulation 
procedure is such that an ensemble of perfect binary trees are generated 
up to a maximum upper bound for the generation $n$ called the depth. 
Afterwards, the edges of the trees are assigned to the state of being open 
with probability $p$ and to the state of being closed with the probability 
$q=1-p$. This yields a random forest of trees within one instance of 
simulation. The tree that contains the root is sieved out as a realisation 
of the Bienaym\'e-Galton-Watson process. Note that, the nodes that are 
lying at the maximum depth are not considered as leaves.
\begin{figure}
	\centering
	\begin{minipage}{.43\textwidth}
		\includegraphics[width=\textwidth]{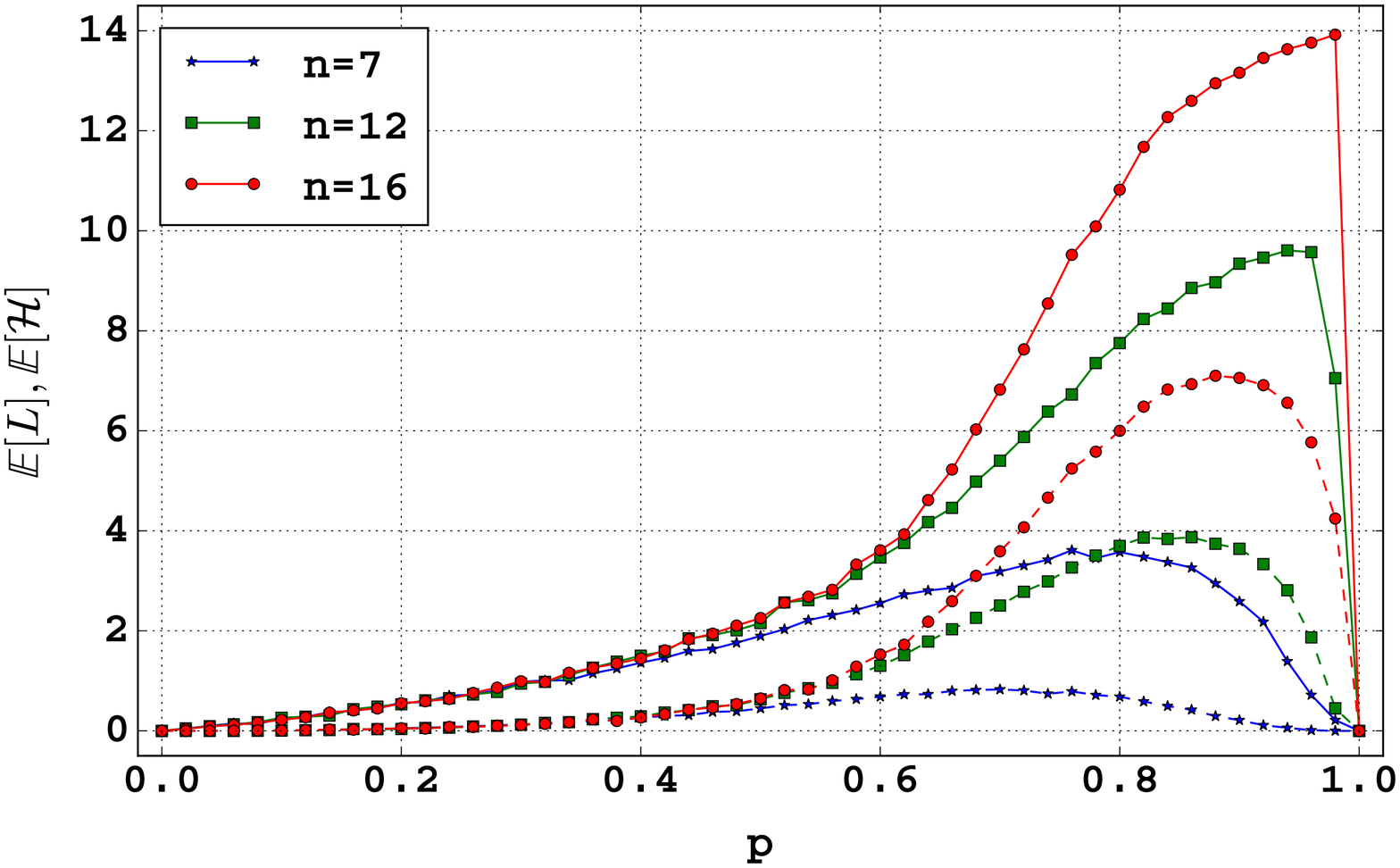}
	\end{minipage}\qquad
	\begin{minipage}{.43\textwidth}
		\includegraphics[width=\textwidth]{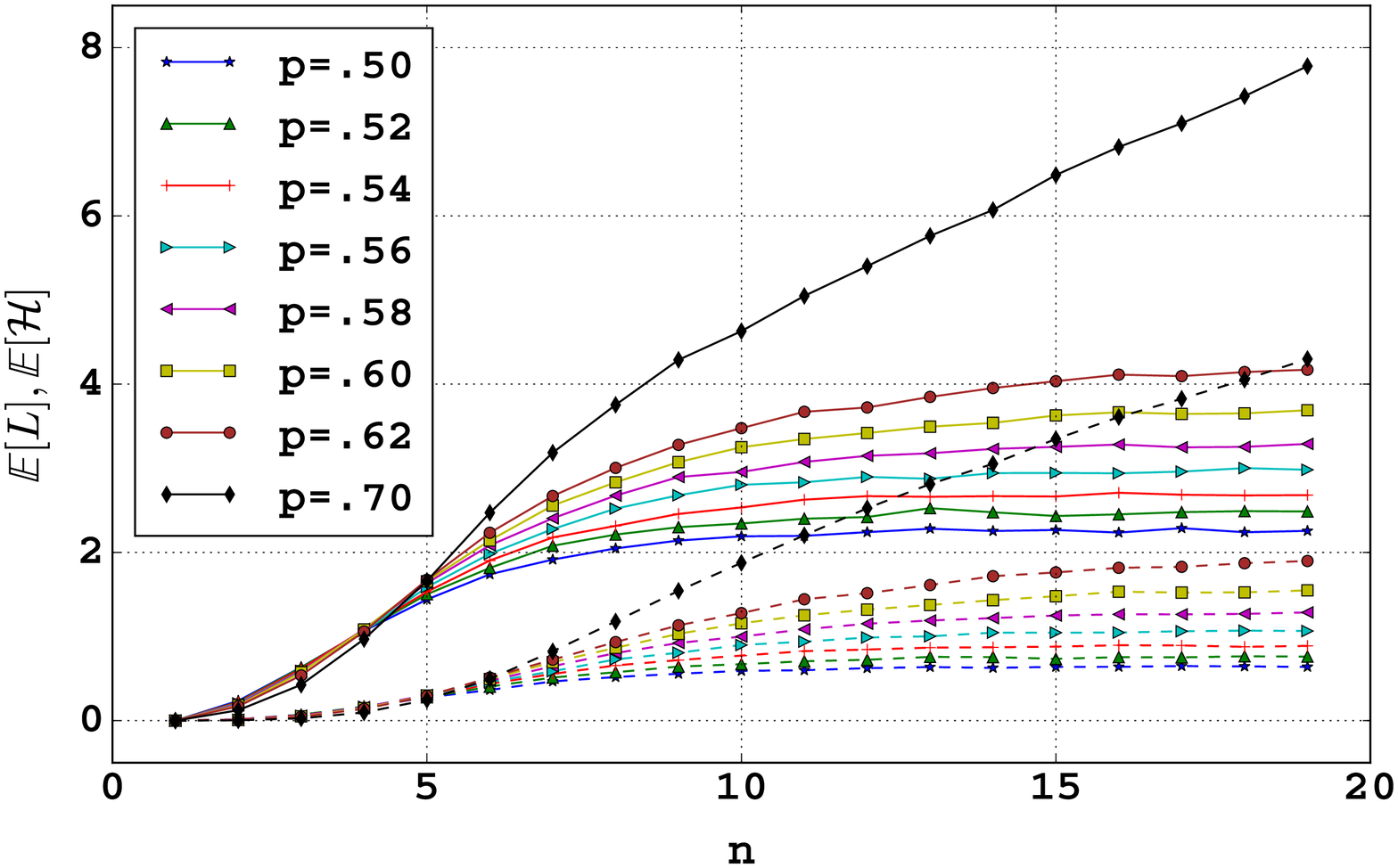}
	\end{minipage}
	\caption{(left) The average codeword length $\mathbb{E}[L]$ 
		 depicted by solid lines and average entropy 
		 $\mathbb{E}[\mathcal{H}]$ depicted by dash lines 
		 versus the percolation density $p$ for three different 
		 depths $n=7,12$ and $16$. Note that when $p=1.0$ both 
		 quantities drop to zero since there are no leaves 
		 present at $p=1.0$. This is expected since there is no 
		 information due to randomness at this density. (right) 
		 The same quantities versus the maximum depth $n$ for 
		 various percolation densities $p$. Again the solid lines 
		 represent the average codeword length and the dashed lines
		 represent the average entropy. It is seen that both 
		 quantities are bounded as long as 
		 $1/2\leq p <1/\sqrt[3]{4}$. For $p=0.7$ observe that they 
		 grow exponentially as $n$ grows. }
	\label{fig:lh_npdependance}
\end{figure}
The simulation results are in agreement with the results yielded above. In
Fig.~\ref{fig:lh_npdependance} on the left pane, the dependence of 
$\mathbb{E}[L]$ and $\mathbb{E}[\mathcal{H}]$ on the maximum generation (depth) 
$n$ and the percolation density $p$ are shown. It is observed that the average 
codeword length and the average entropy versus $p$ monotonically increase until 
they reach their maximum and then reach the value zero when $p=1.0$. This is due
to the fact that when $p=1.0$ there are no leaves that contribute to $L$ 
and $\mathcal{H}$. Moreover, in the same figure on the right pane, the 
dependence of $\mathbb{E}[L]$ and $\mathbb{E}[\mathcal{H}]$ on the depth of
the simulation is depicted. It is clear that so long 
$1/2 \leq p < \sqrt[3]{1/4}$ both quantities remain finite. To make this 
stand out, the average codeword length and the entropy at $p=0.7$ are plotted and 
it is seen that in contrast to the other percolation densities, $\mathbb{E}[L]$ and 
$\mathbb{E}[\mathcal{H}]$ grow exponentially. In Fig.~\ref{fig:avg_length_entropy} 
the average codeword length and the entropy yielded by the simulation results are 
plotted against their analytical expressions given by Eq.~\eqref{eq:avg_entropy} 
and Eq.~\eqref{eq:avg_codewordlength} respectively. The dashed lines are the analytical 
expressions describing the asymptotic behaviour of the aforementioned quantities in 
the limit of $n\rightarrow\infty$. Obviously, there is a gap between the asymptotics 
yielded by the simulation results due to the fact that while deriving the expressions 
Eq.~\eqref{eq:avg_entropy} and Eq.~\eqref{eq:avg_codewordlength} it was assumed that 
$\Lambda(\cC)$ is a constant when $n\rightarrow\infty$ and hence substituted by its 
means. Yet, for the simulation results, the normalisation factor $\Lambda$ is calculated 
exactly. Nevertheless, it is still promising to indeed observe the that the average 
codeword length and the average entropy both remain finite as $n$ increases as long 
as $1/2\leq p < \sqrt[3]{1/4}$.

\section{Conclusion}
It is demonstrated here that a given percolation cluster generated via a 
Bernoulli percolation process on a perfect binary tree can be regarded as 
an encoding scheme with a set of symbols identified as the leaves of the 
cluster, along with their associated probabilities satisfying the Bernoulli 
probability measure. The codewords are simultaneously yielded by traversing 
the open paths from the root of the tree to their respective leaves. It was 
proven that with this very configuration, one can still have a set of infinite 
symbols and keep the amount of the information and the average codeword 
length finite by choosing the percolation density $p$ appropriately.

One of the aims of this paper was to show that a branching process as simple 
as Bienaym\'e-Galton-Watson process could have potential applications in computer 
science. It may have deep implications in how one algorithmically encodes large 
data, stores, compresses and also importantly encrypts them. Other 
non-trivial branching processes may be even found more crucial in shaping our 
understanding of data integrity and storage or encryption methods that are far 
challenging to be breached. Also the author wants to note that the 
Bienaym\'e-Galton-Watson process has already been used to study some natural 
processes. For instance in the electron multiplier detector or nuclear chain 
reactions~\cite{shockley1938,hawkins1944}. Not only limited to these, other 
processes such as the birth-death process which can be utilised to find the 
rate at which a new species emerges~\cite{yule1925}. It may be that the nature 
intrinsically encodes the information in such a manner that the entropy remains 
bounded yet the possible outcomes are infinitely many.

\section*{Acknowledgement}
The author wants to cordially thank Prof Aleksei Chechkin for his 
constructive suggestions and critical views and Prof Stephan Foldes for 
initiating useful discussions that shaped many of the building blocks of 
this article. The author further expresses his gratitudes towards Prof 
Sylvie Roelly for her constructive views and supports.
\newpage
\bibliographystyle{unsrt}  
\bibliography{references}  

%
%

\end{document}